\documentclass{evn2002}
\usepackage{graphicx}
\begin{document}
   \title{Proper motion in Cygnus A}

   \author{U. Bach, T.P. Krichbaum, W. Alef, A. Witzel and J.A. Zensus
          }

   \institute{Max Planck Institut f\"ur Radioastronomie, Auf dem H\"ugel 69, 53121 Bonn, Germany}

   \abstract{Our recent VLBI observations of the prominent FR\,II radio galaxy Cygnus\,A with the EVN and the VLBA reveal a pronounced two-sided jet structure. At 5\,GHz, we now have 4 epochs from 1986, 1991 (Carilli et al., 1991 \& 1994), 1996 and 2002 from which we could derive the kinematics of the jet and counter-jet. On the jet side and on mas scales, the jet seems to accelerate from $\beta_{\rm app}\approx 0.1-0.2$ (Krichbaum et al. 1998) at core-separations near 1\,mas to $\beta_{\rm app}\approx 0.4-0.6$ at $r \geq 4$\,mas
($H_0=100$\,km\,s$^{-1}$\,Mpc$^{-1}$, $q_0=0.5$). For the first time we also measure significant structural variability on the counter-jet side. For this, we derive a motion of $\beta_{\rm app}=0.35\pm0.2$ at $r=9.5$\,mas. The flat spectrum of the inner region of the counter-jet (free-free absorption) and the frequency dependence of the jet to counter-jet ratio support strong evidence for an obscuring torus in front of the counter-jet (Krichbaum et al. 1998).
   }

   \maketitle
%

\section{Introduction}
Cygnus\,A is one of the first and strongest objects detected in the radio (\cite{bolton}). It is the closest ($z=0.057$) strong FR\,II radio galaxy and therefore a key object for detailed studies of FR\,II nuclei. In the radio bands, Cygnus\,A is characterized by two strong lobes separated by $\sim$\,2$^\prime $ in the sky, and two highly collimated jets connecting the lobes with the core (\cite{perley}, \cite{carilli1991}). On VLA scales, the jet is oriented along P.A. $\sim 285^\circ$ and the fainter counter-jet along P.A. $\sim 107^\circ$. Due to the large inclination of the jet with respect to the observer and correspondingly reduced relativistic effects, Cygnus\,A is an ideal candidate for detailed studies of its jet physics, which is thought to be similar to those in the more beamed quasars (e.g. Barthel 1989). A detailed description of the arc second structure of Cygnus\,A is given in Carilli \& Harris (1996) and Carilli \& Barthel (1996).

We present and discuss the preliminary results of our new VLBI images of the core region of Cygnus\,A from an 1.6\,GHz EVN observation (1998.1), two 4.9\,GHz VLBI observations with the EVN (1996.8) and the VLBA (2002.0) and an 8.4\,GHz EVN observation performed simultaneously to the first 4.9\,GHz observation.
%
\section{Observations and data reduction}
Cygnus\,A was observed with the EVN at 4.9\,GHz and 8.4\,GHz in Mk\,III Mode B (28\,MHz bandwidth) in October 1996 and at 1.6\,GHz in Mk\,III Mode A (56\,MHz bandwidth) in February 1998. A second epoch at 4.9\,GHz was observed with the VLBA, a single VLA antenna and Effelsberg with 32\,MHz bandwidth in dual circular polarization in January 2002. Amplitude calibration and fringe-fitting, were carried out in {\sc AIPS} using standard techniques. The T$_{\rm Sys}$ measurements at 1.6\,GHz were dominated by the radio lobes of Cyg\,A which are inside the primary beams of most telescopes at this frequency, and therefore the receivers saturate. In this case the T$_{\rm Sys}$ measurements have to be replaced by the total flux of the source (\cite{anantharamaiah}) which is for Cyg\,A $\sim$4500\,Jy at 1.6\,GHz. Phase- and amplitude self-calibration, as well as the imaging of the source, were done in {\sc Difmap}.
   \begin{figure}
   \centering
   \includegraphics[angle=-90,width=0.47\textwidth]{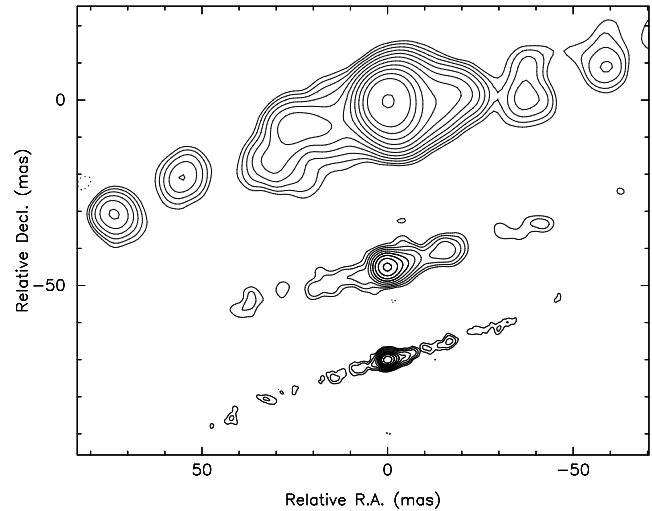}
      \caption{Jet and counter-jet of Cygnus\,A at 1.6\,GHz (top), 4.9\,GHz (middle) and 8.4\,GHz (bottom) from the EVN observations in 1996 and 1998. The maps are convolved with a 10, 3.5 and 2\,mas circular beam respectively, the lowest contours are 10, 2 and 3\,mJy/beam and the brightest peak is 0.73, 0.67 and 1.3\,Jy/beam.
              }
         \label{EVNmaps}
   \end{figure}
   \begin{figure}
   \centering
   \includegraphics[angle=-90,width=0.48\textwidth]{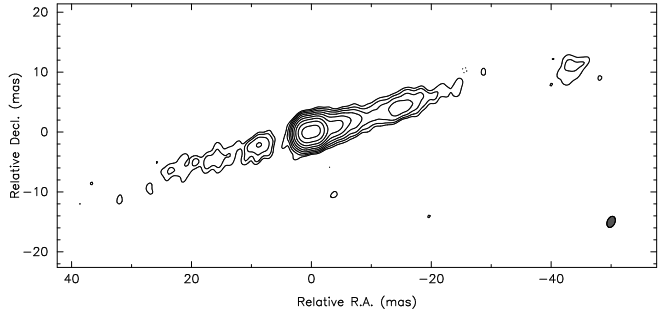}
      \caption{4.9\,GHz VLBA map of the jet and counter-jet of Cygnus A in Jan. 2002. The beam size is $1.95\times1.28$ at $-24.5^\circ$, the lowest contour is 0.5\,mJy/beam and the brightest peak is 0.137\,Jy/beam.
              }
         \label{VLBAmap}
   \end{figure}
\section{Results}
The final maps of the core and of the inner jet and counter-jet of Cygnus\,A are shown in Figures~\ref{EVNmaps}~\&~\ref{VLBAmap}. Only the central part of the 1.6\,GHz EVN map is displayed to make it comparable to the higher frequency maps. The maps show a straight two-sided jet with some prominent features which are seen at every frequency. The EVN observations confirm the observations of Krichbaum et al. (1998) which for the first time showed the counter-jet close to the core of Cygnus\,A. In earlier observations the counter-jet was only seen slightly above the noise level (Carilli et al. 1991 \& 1994). In all of our VLBI maps the western jet is orientated along P.A.$=288^\circ\pm3^\circ$ and the eastern counter-jet along P.A.$=105^\circ\pm3^\circ$. The angles differ slightly ($\sim5^\circ$) from the large-scale structure seen by the VLA and the angle between both jets is not exactly $180^\circ$. This departure from a straight line was also observed in earlier VLBI observations (\cite{carilli_perley})
\subsection{Model-fits}
After imaging in {\sc Difmap} we model fitted circular gaussian components to the self-calibrated data. We chose circular components because of the reduced degrees of freedom and because they are easier to handle in {\sc Difmap} than elliptical components. We expect from the amplitude calibration and from the uncertainties in the model-fits a maximum error of 15\% for the flux density measurements. The FWHM of the components has an error of about 15\%, too, which we derived from the variations between different model-fits. This method was also used for the calculation of the positional errors.
\begin{figure}
   \centering
   \includegraphics[angle=0,width=0.48\textwidth]{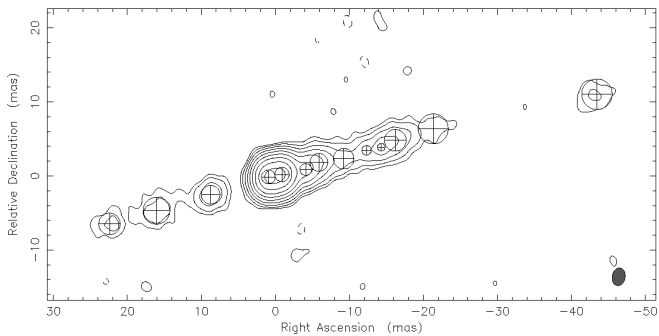}
      \caption{Model-fit to the 4.9\,GHz VLBA map of the jet and counter-jet of Cygnus A in Jan. 2002. The lowest contour is 1\,mJy/beam.
              }
         \label{VLBAmap}
   \end{figure}
\section{Discussion}
\subsection{Identification of components}
To investigate the kinematics and the spectral properties of Cygnus\,A at different frequencies and epochs, we tried to identify individual jet components by their distance from the core, flux density, size and brightness temperature. All of the components identified are listed in Table~\ref{identcomp}. Since the observations were not done in phase-referencing mode, the absolute position information is lost, and it is not possible to tell which components, if any, are stationary. The positions in Table~\ref{identcomp} are measured with respect to the map phase center. Given that the $\tau=1$ surface moves closer to the true core for shorter wavelengths, the position of the brightest, usually innermost peak is likely not to be the same in maps at different wavelengths. From the comparison of profile plots and the shift of components between two simultaneously observed maps at 4.9\,GHz and 8.4\,GHz a value of $\sim 0.5$\,mas could be derived for the core shift. Extrapolating a core shift of $\sim0.1$\,mas between 22\,GHz and 43\,GHz (\cite{krichbaum}) and our measurement we would expect a core shift of 1-2\,mas between 1.6\,GHz and 4.9\,GHz.
\begin{table}[hbtp]
\caption[]{List of all components which could be identified between various frequencies and epochs.}
\label{identcomp}
\begin{minipage}{\linewidth}
\renewcommand{\footnoterule}{}
\scriptsize
\center
\begin{tabular}{crrrc}
\hline\hline
\multicolumn{5}{c}{6\,cm (1996.81)} \\
Comp.\footnote{\scriptsize J and C are identifiers for jet and counter-jet components. S: Flux density, r: distance from the core, PA: Parallactic angle, FWHM: Full Width Half Maximum} & S [Jy] &\multicolumn{1}{c}{r [mas]} & \multicolumn{1}{c}{P.A. [$^\circ$]} & FWHM [mas]\\
\hline
C04 & 0.005 & --32.04 $\pm$0.33 & 110.29 $\pm$0.6 & 3.95\\
C03 & 0.014 & --21.47 $\pm$0.83 & 106.49 $\pm$2.2 & 3.33\\
C02 & 0.009 & --17.30 $\pm$0.67 & 105.56 $\pm$2.2 & 2.68\\
C01 & 0.021 & --8.73 $\pm$0.78 & 105.83 $\pm$5.1 & 3.10\\
J00 & 0.741 & 0 & 0 & 1.94\\
J01 & 0.171 & 2.99 $\pm$0.57 & -80.64 $\pm$7.8 & 2.29\\
J03 & 0.037 & 6.92 $\pm$0.25 & -73.11 $\pm$2.1 & 1.00\\
J04 & 0.019 & 11.74 $\pm$0.52 & -74.30 $\pm$2.5 & 2.09\\
J06 & 0.052 & 16.49 $\pm$0.68 & -74.20 $\pm$2.4 & 2.73\\
J07 & 0.007 & 25.54 $\pm$0.33 & -77.27 $\pm$0.8 & 1.34\\
J08 & 0.009 & 43.91 $\pm$0.61 & -75.03 $\pm$0.8 & 2.43\\
\hline
\multicolumn{5}{c}{6\,cm (2002.03)} \\
\hline
C03 & 0.005 & --23.57 $\pm$0.21 & 106.19 $\pm$0.5 & 2.52\\
C02 & 0.010 & --16.81 $\pm$0.32 & 106.23 $\pm$1.1 & 3.81\\
C01 & 0.014 & --9.07 $\pm$0.21 & 105.50 $\pm$1.3 & 2.51\\
J00a & 0.355 & 0.95 $\pm$0.15 & 97.50 $\pm$9.1 & 1.84\\
J00b & 0.382 & 0.91 $\pm$0.14 & -78.24 $\pm$8.7 & 1.98\\
J01 & 0.037 & 4.25 $\pm$0.12 & -77.77 $\pm$1.6 & 1.39\\
J03 & 0.011 & 9.71 $\pm$0.22 & -76.13 $\pm$1.3 & 2.67\\
J04 & 0.007 & 12.82 $\pm$0.10 & -74.26 $\pm$0.4 & 1.18\\
J06 & 0.014 & 17.59 $\pm$0.22 & -73.01 $\pm$0.7 & 2.62\\
J07 & 0.005 & 22.73 $\pm$0.32 & -73.83 $\pm$0.8 & 3.87\\
J08 & 0.008 & 44.80 $\pm$0.33 & -75.79 $\pm$0.4 & 3.92\\
\hline
\multicolumn{5}{c}{18\,cm (1998.14)} \\
\hline
C04 & 0.042 & --33.86 $\pm$1.06 & 113.75 $\pm$1.8 & 5.19\\
C03 & 0.073 & --21.11 $\pm$0.49 & 110.11 $\pm$1.3 & 2.42\\
J00 & 0.923 & 0 & 0 & 8.11\\
J06 & 0.113 & 18.42 $\pm$1.59 & --88.25 $\pm$4.9 & 7.91\\
J08 & 0.022 & 35.98 $\pm$0.45 & --89.71 $\pm$0.7 & 2.21\\
\hline
\multicolumn{5}{c}{3.6\,cm (1996.81)} \\
\hline
C02 & 0.006 & --14.34 $\pm$0.08 & 109.01 $\pm$0.4 & 0.22\\
C01 & 0.035 & --7.42 $\pm$0.49 & 107.58 $\pm$3.6 & 2.66\\
J00 & 1.557 & 0 & 0 & 1.280\\
J03 & 0.024 & 7.19 $\pm$0.19 & --74.79 $\pm$1.5 & 0.4\\
J04 & 0.007 & 11.07 $\pm$0.08 & --74.84 $\pm$0.4 & 0.96\\
J06 & 0.034 & 16.97 $\pm$0.35 & --74.38 $\pm$1.2 & 1.89\\
\hline
\end{tabular}
\end{minipage}
\end{table}
\subsection{Spectral index}\label{sec:specindx}
\begin{figure}
   \centering
   \includegraphics[angle=-90,width=0.45\textwidth]{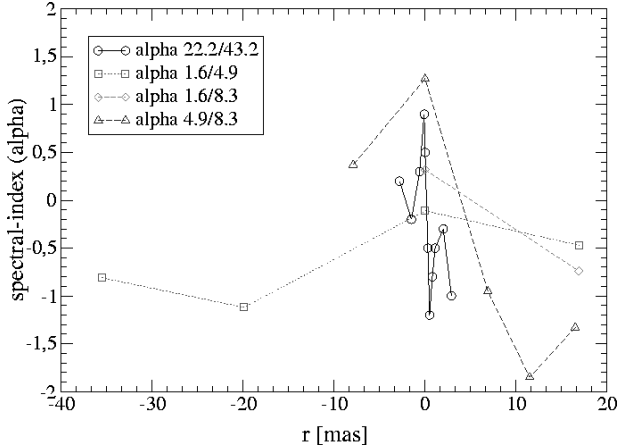}
      \caption{Spectral-indices between 1.6\,GHz and 43\,GHz vs. core-distance. The indices between 22.2\,GHz and 43.2\,GHz are from \cite{krichbaum}.
              }
         \label{specindx}
   \end{figure}

The spectral indices of all jet and counter-jet components, which could be cross-identified at various frequencies, are plotted in Figure~\ref{specindx}. The core spectrum is flat at lower frequencies, and becomes inverted at higher frequencies, which has also been observed between 22\,GHz and 43\,GHz (\cite{krichbaum}). The western jet shows a steep spectrum (--0.3 to --1.5) typical for a region of optically thin synchrotron emission. The very steep components of the outer jet between 4.9\,GHz and 8.4\,GHz are probably due to a reduced flux density of the outer jet at 8.4\,GHz. The outer region of the counter-jet shows a steep spectrum too, but in the inner region ($r<8$\,mas) the spectrum is slightly inverted. Unfortunately, the outer components of the counter-jet could only be measured between 1.6\,GHz (1998.14) and 4.9\,GHz (1996.81), and it cannot be ruled out that their flux density varies over 1.5 years. If the measurements can be trusted which is supported by measurements of Krichbaum et al. (1998), the inverted spectrum of the inner counter-jet could be caused by free-free-absorption by a foreground absorber. UV spectroscopy (Antonucci {\it et al.} 1994) and optical spectro-polarimetry (Ogle {\it et al.} 1997) also show evidence for a hidden broad line region. According to the unified scheme, this is strong evidence for an obscuring torus around the central engine and is consistent with 21\,cm absorption line VLBI (\cite{Blanco}).
\subsection{Jet to counter-jet ratio}
\begin{figure}
   \centering
   \includegraphics[width=0.45\textwidth]{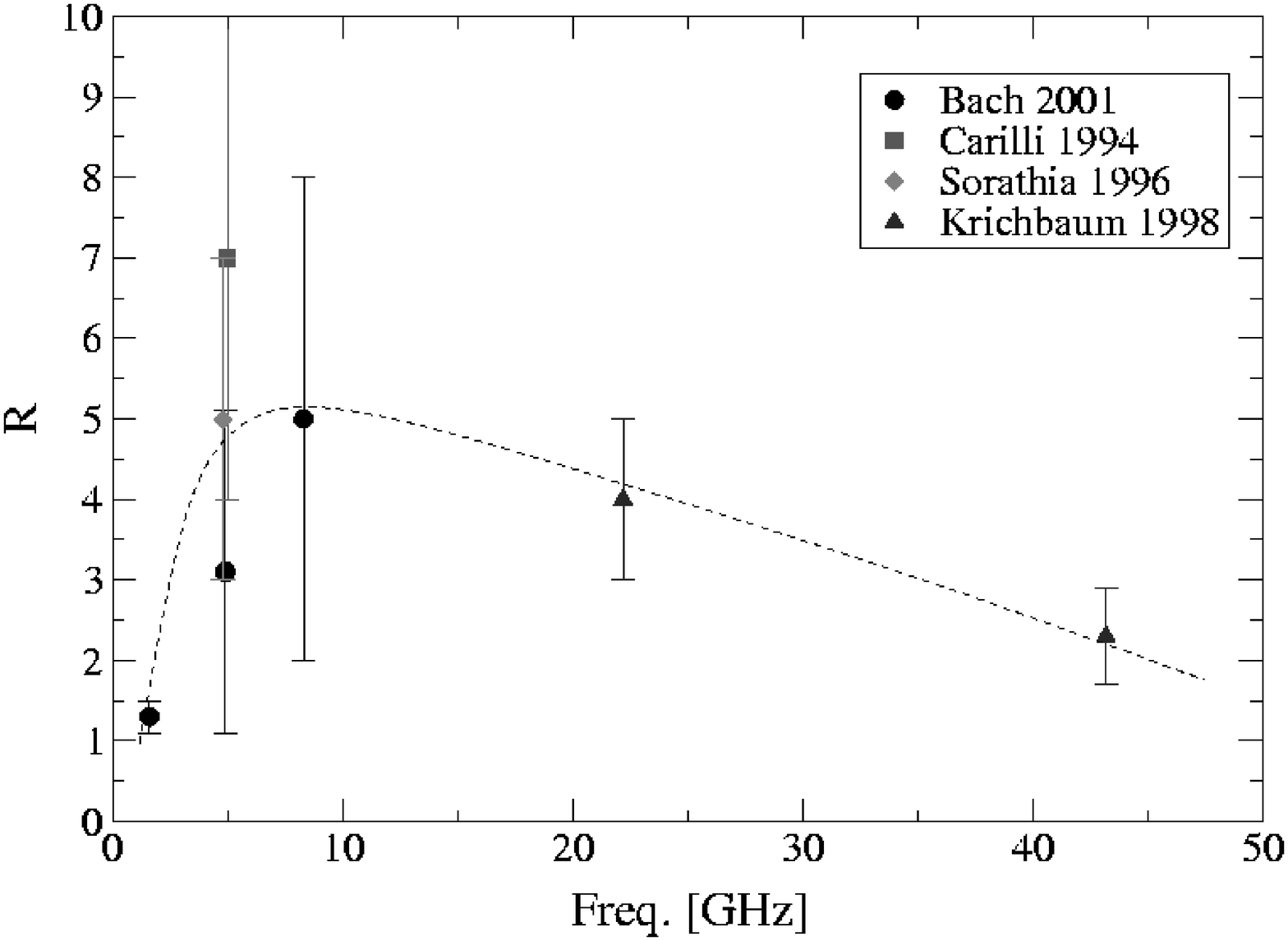}
      \caption{Frequency dependence of the jet to counter-jet ratio R. The dashed line illustrates the trend of the measurements.
              }
         \label{freq-R}
   \end{figure}

The determination of the jet to counter-jet ratio depends on the true core position. Since we don't know exactly were the core is, we derived the jet to counter-jet ratio by numerical integration along one dimensional profiles of the jets for various core positions. In addition, the jet to counter-jet ratio was measured with {\tt TVSTAT} in {\sc AIPS}. The results are presented in Table~\ref{tab:jettocjet}. The ratio of the flux densities of the jet and counter-jet allows, via the apparent velocity of the jet (see Sec.~\ref{beta}), to calculate the inclination of the jet axis with respect to the observer. For this we derive 80$\pm8^\circ$. In Figure~\ref{freq-R} the various measurements are plotted against frequency. Normally, it would be expected that the jet to counter-jet ratio is constant with frequency, but if the assumed obscuring torus (see Sec.~\ref{sec:specindx}), due to its inclination, affects more the inner part of the counter-jet, such a frequency dependence could be observed. At longer wavelengths the absorbing region becomes small compared to the beam size and the effect is diluted, while at shorter wavelengths the torus becomes visible and reduces the flux density of the counter-jet, which will cause a rise of the jet to counter-jet ratio. Towards mm-wavelengths, the absorber would become optical thin and the jet to counter-jet ratio will decrease. Since the obscuring torus seems not to affect the jet to counter-jet ratio at 1.6\,GHz we presume an upper limit of 10\,mas (corresponding to $9\,h^{-1}$\,pc) for the projected size of the torus.
\begin{table}[hbtp]
\center
\footnotesize
\caption[]{Jet to counter-Jet ratio at different core positions and frequencies}
\label{tab:jettocjet}
\centering
\begin{tabular}{lcc}
\hline\hline
$\lambda$ [cm] & R$_{\rm min}$ & R$_{\rm max}$\\
\hline
18 & 1.2 & 1.5\\
6 & 1.3 & 3.1\\
3.6 & 1.2 & 5.8\\
\hline
\end{tabular}
\end{table}
\subsection{Kinematics of the jet components}\label{beta}
It was possible to cross-identify 7 jet components between the two epochs at 4.9\,GHz  and to follow their trajectories. Our findings are in agreement with previous observations from 1986.89 and 1991.44 by Carilli et al. (1991 \& 1994). In Table~\ref{positions} all components are listed in which motion is detected. Counter-jet component C01 is only poorly visible in the 1986 map and Carilli did not believe this component to be real, while our observations confirm the existence of C01.
\begin{table}[hbtp]
\center
\caption{Position of jet and counter-jet components from 1986 to 2002.}
\label{positions}
\scriptsize
\begin{minipage}{\linewidth}
\renewcommand{\footnoterule}{}
\begin{tabular}{lrrrrrrrr}
\hline\hline
 & 1986.89\footnote{\scriptsize From Carilli et al. (1991)} & 1991.44\footnote{\scriptsize From Carilli et al. (1994)}& 1996.85 & 2002.03\\
Comp. & r [mas] & r [mas] & r [mas] & r [mas] \\
\hline
C03 & & & --21.47 $\pm$0.83 & --23.57 $\pm$0.21\\
C01 & --7.30 $\pm$0.20 &  & --8.73 $\pm$0.78 & --9.07 $\pm$0.21\\
J01 & & & 2.99 $\pm$0.57 & 4.25 $\pm$0.12\\
J03 & 4.540$\pm$0.05& 4.94 $\pm$0.06& 6.92 $\pm$0.25 & 9.71 $\pm$0.22\\
J04 & & & 11.74 $\pm$0.52 & 12.82 $\pm$0.10\\
J06 & 14.68 $\pm$0.32 & 15.53 $\pm$0.16 & 16.49 $\pm$0.68 & 17.59 $\pm$0.22\\
J08 & & & 43.91 $\pm$0.61 & 44.80 $\pm$0.33\\
\hline 
\end{tabular}
\end{minipage}
\end{table}

The calculated apparent velocities $\beta_{\rm app}$ are listed in Table~\ref{tab:appbeta} ($H_0=100\,h$\,km\,s$^{-1}$\,Mpc$^{-1}$ and $q_0=0.5$). Component J06 shows a constant apparent velocity of 0.5\,$c$ over the last 16 years. J01 and J08 have a comparable $\beta_{\rm app}$ to J06. J08 was also confirmed by Sorathia et al. (1996). J03 seems to have accelerated from 0.2\,$c$ to 1.3\,$c$ which is very unlikely. The high apparent speed of 1.3\,$c$ cannot be expected for a relativistic jet with a inclination of $\sim80^\circ$ to the line of sight, it is rather due to a false component identification. The counter-jet component C01 which had a significant $\beta_{\rm app}$ of 0.35\,$\pm$0.2\,$c$ in 1996 has slowed down to a no more significant speed of 0.16\,$\pm$0,39\,$c$. Due to its large FWHM and the low flux density, also the apparent velocity of C03 should be handled with care. 

\begin{table}[hbtp]
\center
\caption{Apparent $\beta$ of jet and counter-jet components.}
\label{tab:appbeta}
\scriptsize
\begin{tabular}{lrrrr}
\hline
Comp. & $\beta_{\rm app\,86/91}$ & $\beta_{\rm app\,86/96}$ & $\beta_{\rm app\,91/96}$ & $\beta_{\rm app\,96/02}$ \\
\hline
C03 & & & & 1.01 $\pm$0.41\\
C01 & & 0.35 $\pm$0.20 & & 0.16 $\pm$0.39\\
J01 & & & & 0,61 $\pm$0,28\\
J03 & 0.22 $\pm$0.04 & 0.60 $\pm$0.08 & 0.92 $\pm$0.14 & 1.35 $\pm$0.16\\
J06 & 0.47 $\pm$0.20 & 0.46 $\pm$0.12 & 0.44 $\pm$0.18 & 0.53 $\pm$0.34\\
J08 & & & & 0.43 $\pm$0.33\\
\hline
\end{tabular}
\end{table}

\begin{figure}
   \centering
   \includegraphics[width=0.55\textwidth]{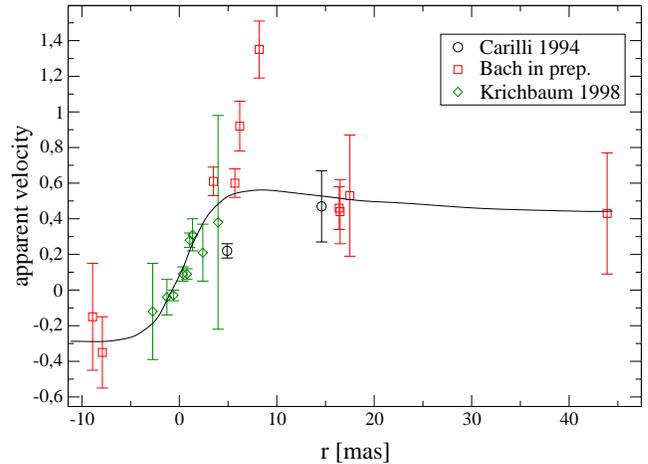}
      \caption{Collection of all so far measured apparent velocities plotted vs. their core distance. The line illustrates the trend of the measurements.
              }
         \label{betaapp}
   \end{figure}

In Figure~\ref{betaapp} all measured apparent speeds are plotted against their core distance. The measurements in the inner core region are from Krichbaum et al.\,(1998). The jet and the counter-jet seem to accelerated from sub-pc to pc scales and the jet seems to be faster than the counter-jet.

\section{Conclusion}
Although our results are still preliminary, the spectral properties give a detailed view on the physics at the core of Cygnus\,A. We conclude that the small jet to counter-jet ratio and correspondingly the large inclination with respect to the observer, which reduces relativistic effects, contradict the observed difference in the jet and counter-jet apparent velocities and question the applicability of simple jet-models, in which straight and intrinsically symmetrical jets are assumed. In the near future we will have two more observations spaced by only 6 months and we also proposed phase-referencing observation to solve the question of the location of the true nucleus.

\begin{acknowledgements}
The European VLBI Network is a joint facility of European, Chinese and other radio astronomy institutes funded by their national research councils. The VLBA is an instrument of the National Radio Astronomy Observatory, a facility of the National Science Foundation, operated under cooperative agreement by Associated Universities, Inc. U.B. acknowledges partial support from the EC ICN RadioNET (Contract No. HPRI-CT1999-40003).
\end{acknowledgements}

\end{document}